\documentclass[3p,times,twocolumn]{elsarticle}
 \biboptions{comma,sort&compress}
 
\usepackage{graphicx}
\usepackage{here}
\usepackage{ecrc}

\volume{00}

\firstpage{1}
\usepackage{subfigure}
\journalname{Nuclear and Particle Physics Proceedings}

\runauth{}
\usepackage{amsmath}

\jid{nppp}

\jnltitlelogo{Nuclear and Particle Physics Proceedings}

\usepackage{amssymb}

\usepackage[figuresright]{rotating}

\def\beq{\begin{equation}}
\def\eeq{\end{equation}}
\def\bea{\begin{eqnarray}}
\def\eea{\end{eqnarray}}

\def\bibi{\bibitem}
\def\a{\alpha}

\def\P{\Pi}
\def\r{\rho}    
\def\p{\pi}
\def\cd{{\cal D}}
\def\d{\delta}
\def\g{\gamma}
\def\b{\beta}

\def\m{\mu}
\def\n{\nu}

\def\nn{\nonumber}

\begin{document}

\begin{frontmatter}

\title{$\alpha_s$ from the updated ALEPH data for hadronic $\tau$ decays\tnoteref{Conf}}
 \tnotetext[Conf]{Talk given at 18th International Conference in Quantum Chromodynamics
(QCD 15, 30th anniversary), 29 june -- 3 july 2015, Montpellier (France).}

 \author[label1,label1b]{Diogo Boito\corref{cor1}}
\cortext[cor1]{Speaker}
\ead{boito@ifsc.usp.br}

\author[label2]{Maarten~Golterman}
 
\author[label3,label4]{Kim~Maltman}

\author[label5]{James~Osborne}
 
\author[label6]{Santiago~Peris}

 \address[label1]{Instituto de F\'isica de S\~ao Carlos, Universidade de S\~ao Paulo,
CP 369, 13560-970, S\~ao Carlos SP, Brazil}
 
 \address[label1b]{Instituto de F\'isica, Universidade de S\~ao Paulo, Rua do Mat\~ao Travessa R, 187, 05508-090,  S\~ao Paulo, SP, Brazil}
 
 \address[label2]{Department of Physics and Astronomy,
San Francisco State University, San Francisco, CA 94132, US}

 \address[label3]{Department of Mathematics and Statistics, 
York University,  Toronto, ON Canada M3J~1P3}

 \address[label4]{CSSM, University of Adelaide, Adelaide, SA~5005 Australia}

 \address[label5]{Physics Department, University of Wisconsin-Madison, 1150 University Avenue, Madison,
Wisconsin 53706, USA}

\address[label6]{Department of Physics, Universitat Aut\`onoma de Barcelona, E-08193 Bellaterra, Barcelona, Spain}

\pagestyle{myheadings}
\markright{ }
\begin{abstract}
We  extract the strong coupling $\alpha_s(m_\tau^2)$ from the recently updated ALEPH non-strange spectral functions obtained from  hadronic $\tau$ decays.
We  apply a self-consistent analysis method, first tested in  the  analysis of OPAL data, to extract $\alpha_s(m_\tau^2)$ and non-perturbative contributions. 
The analysis yields $\alpha_s^{\rm FO}(m_\tau^2)=0.296\pm0.010 $, using  Fixed Order Perturbation Theory (FOPT), and $\alpha^{\rm CI}_s(m_\tau^2)= 0.310\pm0.014$, using Contour Improved Perturbation Theory (CIPT).  The weighted average of these results with those previously obtained from OPAL data give $\alpha_s^{\rm FO}(m_\tau^2)=0.303\pm 0.009$
and $\alpha_s^{\rm CI}(m_\tau^2)=0.319\pm 0.012$,  which gives, after evolution to the $Z$ boson mass scale,  $\alpha^{\rm FO}_s(m_Z^2)=0.1165\pm0.0012 $ and $\alpha_s^{\rm CI}(m_Z^2)=0.1185\pm0.0015 $, respectively.  We observe that non-perturbative effects limit the accuracy with which $\alpha_s$ can be 
extracted from $\tau$ decay data. 
\end{abstract}

\begin{keyword}  $\alpha_s$, $\tau$ decays, duality violations
\end{keyword}

\end{frontmatter}

\section{Introduction}

The extraction of $\alpha_s$ from hadronic $\tau$ decays 
represents an important test of the evolution of the strong coupling as predicted by the
QCD $\beta$-function.  At and around the $\tau$ mass, $m_\tau\approx 1.78$~GeV, perturbative
QCD can still be used, but realistic analyses  must include the  contribution from non-perturbative effects. 
The standard framework to describe hadronic $\tau$ decays is to organize the QCD description 
in an  operator product expansion (OPE) where, apart from the perturbative contribution and quark-mass corrections,
 the QCD condensates intervene~\cite{BNP}.

Observables such as $R_\tau$,
\beq\label{RtauDef}
R_\tau =\frac{\Gamma \left[ \tau^- \to \nu_\tau {\rm hadrons}(\gamma)  \right]}{\Gamma \left[ \tau^- \to \nu_\tau e^-\bar \nu_e(\gamma)  \right]},
\eeq
can be written as   weighted integrals over the experimentally accessible QCD  spectral functions. The spectral functions have been determined at LEP by the ALEPH~\cite{ALEPH} and OPAL collaborations~\cite{OPAL}. In the specific case of $R_\tau$,
the weight function is that determined by $\tau$-decay kinematics and the integral runs over the total energy of the hadronic system in the final state, $s$, from zero 
to $m_\tau^2$. Since the OPE description is not valid at low-energies the evaluation of the theoretical counterpart is performed exploiting the
analytical properties of the QCD correlators. One writes then a finite-energy sum rule (FESR) where the theoretical  counterpart of the observable 
is obtained from an integral along a complex circle of fixed radius $|s|=m_\tau^2$. In fact, any analytic weight function gives rise to a valid sum rule, and it has
become customary to exploit this freedom in order to analyse several FESRs simultaneously. This type of combined analysis allows for the extraction of $\alpha_s$ as well as  non-perturbative contributions.

The computation, in 2008, of the NNNLO term, $\mathcal{O}(\alpha_s^4)$, of the perturbative expansion of the QCD correlators~\cite{BCK08} triggered several
reanalyses of $\alpha_s$ from $\tau$ decays~\cite{MY08,Davieretal08,alphas2011,alphas2012}. In the process, it was discovered that the correlation matrices 
of the then publicly available spectral functions from
the ALEPH collaboration had a missing contribution from the unfolding procedure~\cite{Manchester}. (It was for this reason that we restricted our attention  to OPAL data in the analyses of Refs.~\cite{alphas2011,alphas2012}.) Recently, 
a new analysis of the ALEPH data became available, employing a new unfolding method, which corrects for this problem in the correlation matrices~\cite{Davieretal14,SpecFunc}. The ALEPH  spectral functions have smaller errors than OPAL's and have the potential to constrain the theoretical description better.  The new set of ALEPH spectral functions motivates the present reanalysis.  

On the theoretical side, two different aspects have received attention recently.  The first one regards the use of the renormalization group in 
improving the perturbative series. There are several prescriptions as to how one  should set the renormalization scale. The most commonly used are
Contour Improved Perturbation Theory (CIPT)~\cite{CIPT} and Fixed Order Perturbation Theory (FOPT)~\cite{MJ05}.  Different prescriptions
lead to different results with the available terms of the perturbative expansion, and therefore to different values of $\alpha_s$.  This discrepancy remains
one of the largest sources of uncertainty in $\alpha_s$ extractions from $\tau$ decays.  Strong evidence in favour of the FOPT prescription has been
given in Refs.~\cite{BJ08,BBJ13} but the issue is still under debate. Here, we chose to perform our analysis using both prescriptions and hence quote two values of $\alpha_s$. 

The second point that has been studied recently is the description of non-perturbative effects.  Since the work of Ref.~\cite{MY08} it is known
that the OPE parameters obtained in some of the recent analyses are inconsistent.  With these parameters one cannot account properly for the experimental results
 when $s_0$, the upper limit of the integration in the FESR, is lowered below $m_\tau^2$.  A strategy that allows for a self-consistent analysis
is the inclusion of Duality Violation (DV) effects in the theoretical description. It is well known that in the vicinity of the Minkowski axis the OPE alone cannot account 
for all non-perturbative effects. In the past, in the description of $R_\tau$, this contribution was systematically ignored due to the fortuitous double zero of the weight function  at the Minkowski axis. In the same spirit, combined analyses of several FESRs were restricted to the so-called {\sl pinched moments}, i.e., moments that have a zero at the the Minkowski axis. Until recently, DVs were not tackled directly and their contribution to final results and errors were not systematically assessed.   

Recent progress in modeling the  DV contribution~\cite{BlokShifmanetal,CGP05,CGP08,CGP09,MJ11} has allowed  for 
analyses that include them explicitly in the FESRs. 
In Refs.~\cite{alphas2011} a new analysis method taking into account DVs explicitly was presented.  This led to a determination
of $\alpha_s$ from OPAL data in a self-consistent way, together with the OPE contribution and DV parameters~\cite{alphas2012}.
In a recent work, we applied the same analysis method to the updated version of the ALEPH non-strange spectral functions~\cite{alphas2015}.

In  the remainder we discuss the main results of Ref.~\cite{alphas2015}.

\section{Analysis framework}

For the sake of self-consistency, here we make a brief review of the framework of our analysis. The details can be found in the original publications Refs.~\cite{alphas2011,alphas2012,alphas2015}.

Fits performed in our analysis  are based on
FESRs of the following  form \cite{Shankaretal,BNP}
\bea
\label{FESR}
I^{(w)}_{V/A}(s_0)&\equiv&\int_0^{s_0}\frac{ds}{s_0}\;w(s)\;\r^{(1+0)}_{V/A}(s)\nn
\\&=&-\frac{1}{2\p i}\oint_{|s|=s_0}\frac{ds}{s_0}\;w(s)\;\P^{(1+0)}_{V/A}(s)\ ,
\eea
where the weight-functions $w(s)$ are  polynomials in $s$, and 
$\P^{(1+0)}_{V/A}(s)$ is given by
\begin{align}
\label{corr}
&i\int d^4x\,e^{iqx}\,\langle 0|T\left\{J_\m(x)J_\n^\dagger(0)\right\}|0\rangle
\nn\\ &=\left(q_\m q_\n-q^2 g_{\m\n}\right)\P^{(1+0)}(s)+q^2 g_{\m\n}\P^{(0)}(s)\ ,
\end{align}
with $s=q^2=-Q^2$, and $J_\m$  one of the non-strange $V$ or $A$ currents
$\bar u\g_\m d$ or $\bar u\g_\m\g_5 d$. The superscripts $(0)$ and $(1)$ refer to
spin. In the sum rule Eq.~(\ref{FESR}), $\rho^{(1+0)}_{V/A}$ is the
experimentally accessible spectral function.  We construct FESRs at several values of $s_0\leq m_\tau^2$ for a given weight function. 

The correlators  $\P^{(1+0)}_{V/A}(s)$ can be decomposed exactly into three
parts
\begin{equation}
\label{th}
\P^{(1+0)}(s)=\P^{(1+0)}_{ \rm pert}(s)+\P^{(1+0)}_{\rm OPE}(s)+\P^{(1+0)}_{\rm DV}(s),
\end{equation}
where ``pert'' denotes perturbative, ``OPE'' refers to OPE corrections   of dimension larger than zero (including quark-mass corrections),  and ``DV'' denotes
the DV contributions to $\P^{(1+0)}(s)$.

It is convenient to write the perturbative contribution   in terms of the physical Adler function~\cite{MJ05,BJ08}, that satisfies 
a homogeneous renormalization group equation.   When treating the
contour integration, one must adopt a prescription for the renormalization scale.  As discussed above,
we perform our analysis within CIPT and FOPT and quote both results. 

The leading contributions from higher dimensions in the OPE 
can be parametrized  with effective coefficients $C_{D}$ as
\begin{equation}
\label{OPE}
\P^{(1+0)}_{\rm OPE}(s)=\sum_{k=1}^\infty\frac{C_{2k}(s)}{(-s)^k}\ .
\end{equation}
The dimension-two quark-mass corrections can safely be neglected for the non-strange correlators.\footnote{We have checked that explicitly.}   Therefore,  in our analysis $C_2=0$.\footnote{An alternative view on the dimension 2 contribution can be found in Refs.~\cite{NZ09,SN09}} The first non-negligible contribution is then  $C_4$, that can be related to the gluon condensate.  However, the weight functions employed in our analysis are polynomials constructed from  combinations of the unity,   $s^2$, and $s^3$. Therefore, in our FESRs, the leading contributions from the OPE arise solely from $C_6$ and $C_8$.  We neglect subleading logarithmic corrections and treat all $C_D$ as constants.  

The DV contribution to the sum rules can be written~as 
\begin{equation}
\label{FESRDV}
\cd_w(s_0)=-\int_{s_0}^\infty \frac{ds}{s_0}\,w(s)\,\r^{\rm DV}(s),
\end{equation}
where $\r^{\rm DV}(s)$ is the DV part of the spectral function in a given channel 
\beq
\label{specDV}
\r^{\rm DV}(s)=\frac{1}{\p}\,\mbox{Im}\,\P^{(1+0)}_{\rm DV}(s),
\end{equation}
that, for $s$ large enough,  we parametrize with the Ansatz of 
Ref.~\cite{CGP05,CGP08}
\begin{equation}
\label{DVpar}
\r^{\rm DV}_{V/A}(s)=\mbox{exp}\left(-\d_{V/A}-\g_{V/A}s\right)\sin\left(\a_{V/A}+\b_{V/A}s\right) .
\end{equation}
This adds 4 new parameters in each channel.

We do not restrict our analysis to pinched weight functions. Rather, we include a weight function that is not pinched
in order to constrain the DV parameters better. We work with three weight functions, namely,
\bea
w_1&=&1, \nn\\ 
w_2&=&1-x^2, \nn \\
w_3&=& (1-x^2)(1+2x),   
\label{moments}
\eea
where $x\equiv s/s_0$. (The weight function $w_3$ is the one determined by kinematics and that yields $R_\tau$.)
This choice is motivated by the fact that we want to perform a self-consistent analysis, including all leading order contribuitons  in the OPE --- without truncating this series arbitrarily. The  extensive explorations of Refs.~\cite{alphas2011,alphas2012} have shown that this set of weight functions fulfills these requirements and allows for a good determination of $\alpha_s$. We remark that these
weight functions also have good perturbative behaviour, in the sense of the analysis performed in~\cite{BBJ13}.

\section{Fits}

We have performed several different fits, combining subsets of the moments of Eq.~(\ref{moments}), fitting to the $V$ or the combined $V$ and $A$ channels. In the fits we include many different values of $s_0$ that lie inside a window $[s_{\rm min}, s_{\rm max}]$ in which  our treatment of the perturbative series, the OPE contributions, and the DVs give an accurate description of the QCD correlator.  One has to vary the value of $s_{\rm min}$ in the fits to check for stability. 
Fits to moments of a single weight function are performed minimizing a standard $\chi^2$. Fits that involve moments of more than one weight function, on the other hand, have too strong correlations to allow for a fit of this type. In this case, one must resort to other measures of fit quality and change the error propagation accordingly in order to account for the strong correlations. The procedure we adopt is discussed in detail in Appendix of Ref.~\cite{alphas2011}.  

We performed a number of consistency tests to assess the robustness of the outcome of our fits.  To corroborate
the results, we performed a study of the posterior probability with  a Markov-Chain Monte Carlo. The results
were also checked for consistency comparing fits to  $V$ and combined analyses of $V$ and $A$ channels.
Another important test is the stability against variations in $s_0$, the upper limit of integration in Eq.~(\ref{FESR}).
The description of moments of the spectral functions must be valid for $s\leq m_\tau^2$. We have checked this stability
for several weight functions. Finally, the outcome of our fits was checked for consistency using the Weinberg sum rules (see Sec. \ref{sec:WSR}).

\section{Main results}

\subsection{Results for $\alpha_s$}
\label{alphas}

The results for $\alpha_s$ that we obtain from   the different fit set-ups are consistent within statistical errors. We choose to quote
as our final value the one obtained from a fit to the  $V$   channel combining moments of the three weight functions $w_1$, $w_2$, and $w_3$,
 of Eq.~(\ref{moments}).  Fits including the $A$ channel  require an extra assumption, namely, 
 that the asymptotic regime assumed in Eq.~(\ref{DVpar}) has already been reached although one works on the tail of the $a_1$ resonance.  
 Although the consistency of the results indicate that this assumption may be fulfilled we prefer to rely on results from $V$ channel only when quoting our final $\alpha_s$ values.  Our final $\alpha_s(m_\tau^2)$ values for $N_f=3$ within the $\overline{\rm MS}$ scheme are
  \begin{align}
& \alpha_s(m_\tau^2) = 0.296 \pm 0.010 \qquad ({\rm ALEPH,\, FOPT}), \nn \\
 &\alpha_s(m_\tau^2) = 0.310 \pm 0.014 \qquad ({\rm ALEPH,\, CIPT}). \nn 
   \end{align}
 The errors  are dominated by statistics, but they also include an estimate of the error due to varying the $s_0$ window and the error due
 to truncation of the perturbative series. When evolved to $m_Z^2$ these results read
  \begin{align}
 &\alpha_s(m_Z^2) = 0.1155 \pm 0.0014 \qquad ({\rm ALEPH, \, FOPT}), \nn \\
 &\alpha_s(m_Z^2) = 0.1174 \pm 0.0019 \qquad ({\rm ALEPH, \, CIPT}) \nn .
 \end{align}
 
 These values are compatible with the ones obtained from the OPAL data~\cite{alphas2012}. Since the data
 sets are independent, the $\alpha_s$ values are virtually uncorrelated. This allows for a weighted average to
 be performed. The averaged values are slightly higher, since values from OPAL data tend to be larger. We find
  \begin{align}
 &\alpha_s(m_\tau^2) = 0.303 \pm 0.009 \,\,\,\, (\mbox{ALEPH and OPAL, FOPT}), \nn \\
& \alpha_s(m_\tau^2) = 0.319 \pm 0.012 \,\,\,\, (\mbox{ALEPH and OPAL, CIPT}), \nn 
 \end{align}
 and at the $Z$ boson mass one has
\begin{align}
 &\alpha_s(m_Z^2) = 0.1165 \pm 0.0012 \,\,\,\, (\mbox{ALEPH and OPAL, FOPT}), \nn \\
& \alpha_s(m_Z^2) = 0.1185 \pm 0.0015 \,\,\,\, (\mbox{ALEPH and OPAL, CIPT}). \nn
 \end{align}

\subsection{Weinberg sum rules}
\label{sec:WSR}

\begin{figure}[ht]
\begin{center}
\subfigure[First Weinberg sum rule, Eq.~(\ref{WSR1}).]{\includegraphics[width=.9\columnwidth,angle=0]{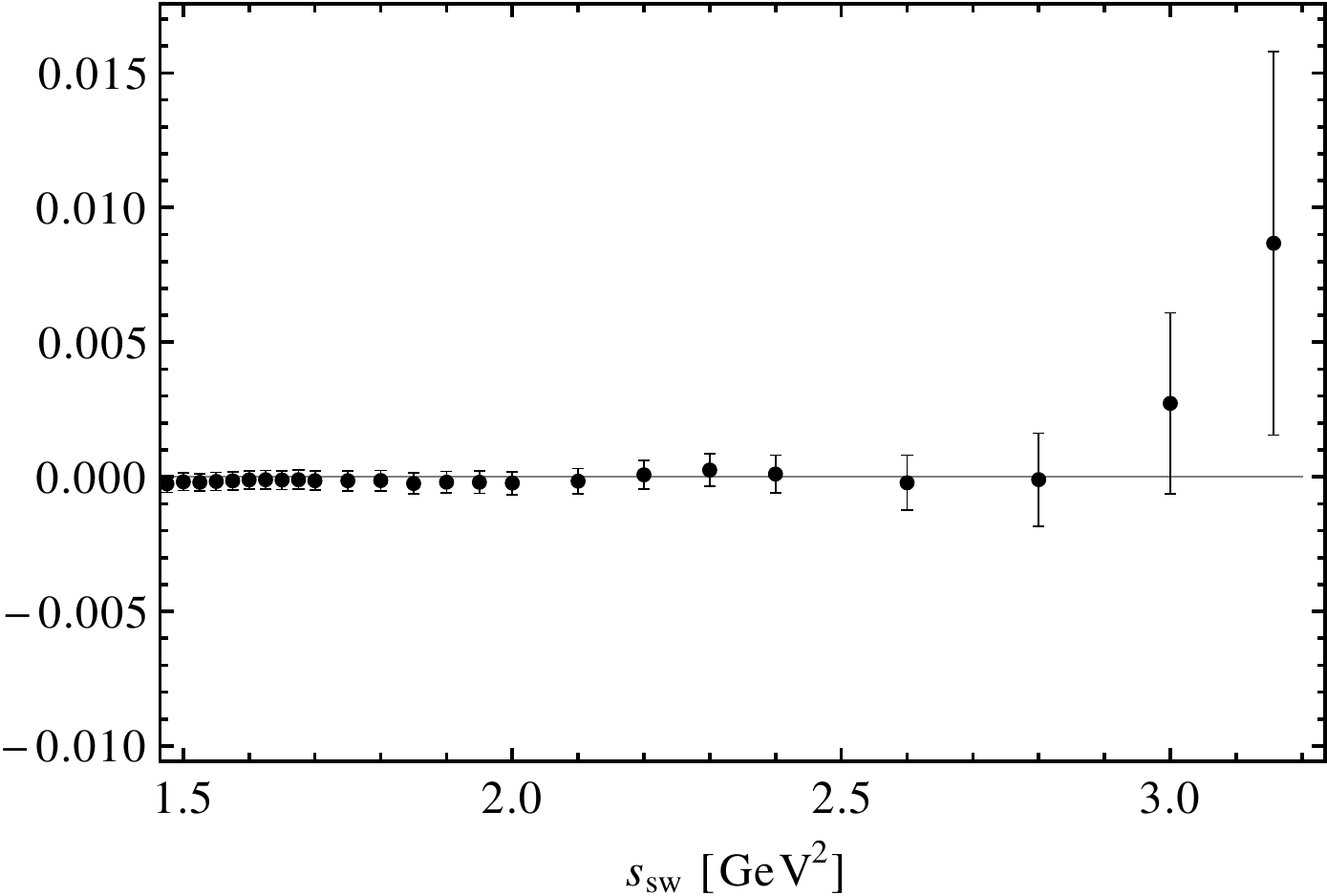}}
\subfigure[Second Weinberg sum rule, Eq.~(\ref{WSR2}).]{\includegraphics[width=.9\columnwidth,angle=0]{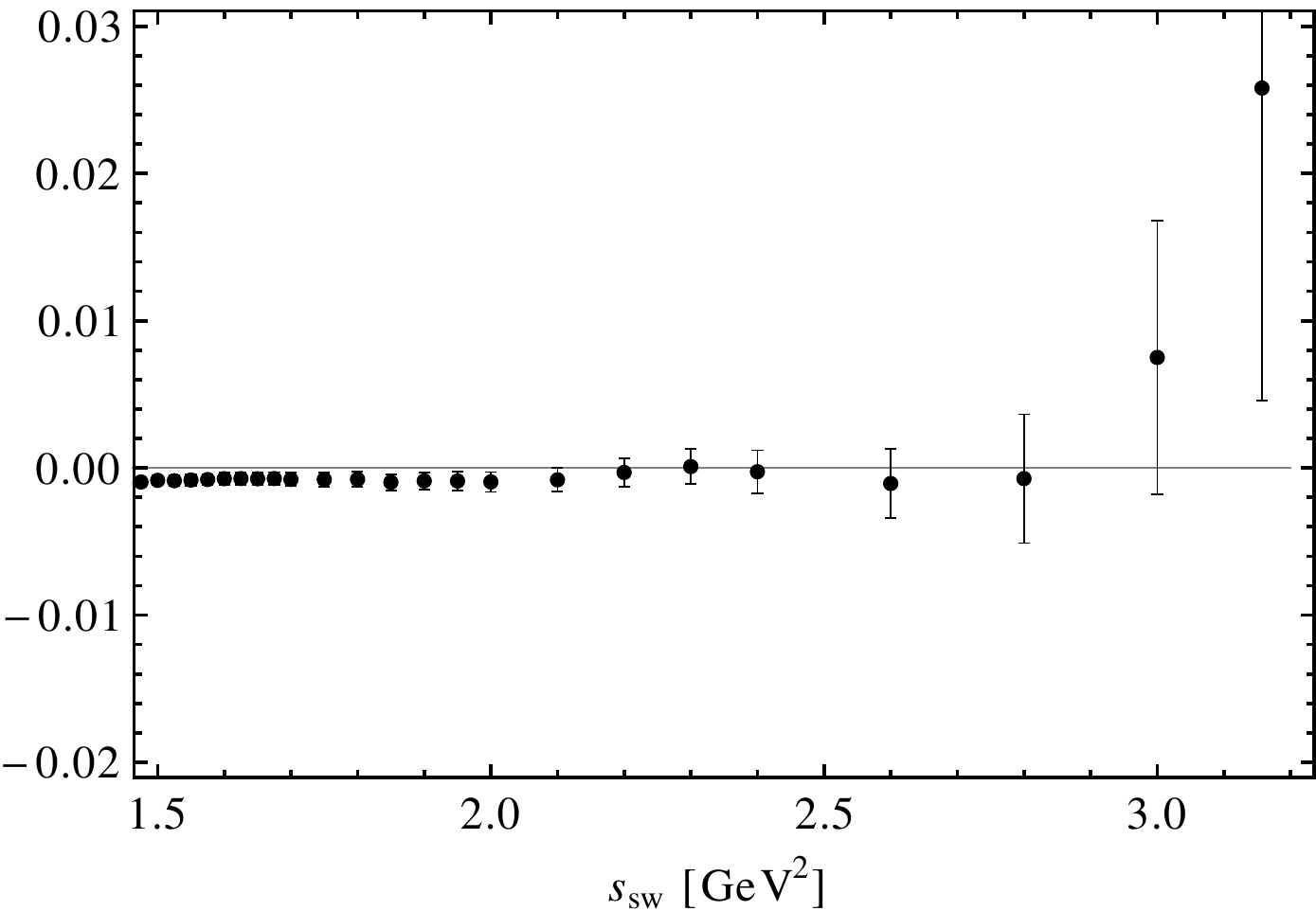}}
\caption{Weinberg sum rules, Eqs.~(\ref{WSR1}) and ~(\ref{WSR2}). Data is used for $s<s_{\rm sw}$ and the DV Ansatz of Eq.~(\ref{DVpar}) is used for $s>s_{\rm sw}$using results from a fit to $V$ and $A$ with moments of $w_1$, $w_2$, and $w_3$, within CIPT. The parameters are given in Tab.~V of Ref.~\cite{alphas2015}.}\vspace{-0.8cm}
\label{WSRfig}
\end{center}     
\end{figure}

Results of combined fits to sum rules of the $V$ and $A$ channels allow for tests of the Weinberg sum rules. 
The first and second Weinberg sum rules  can be written as
\bea
&&\int_0^{\infty} ds\, \left(\rho_V^{(1)} - \rho_A^{(1)} \right) - 2f_\pi^2 = 0,  \label{WSR1}\\
&&\int_0^{\infty} ds\, s\, \left(\rho_V^{(1)} - \rho_A^{(1)}\right) - 2m_\pi^2\,f_\pi^2 = 0,  \label{WSR2}
\eea
where the pion pole contribution has been separated.  For the second sum rule we assumed
that terms of order $m_i\,m_j$, with $i,j=u,d$, can be neglected. 
To check such sum rules, one defines a point $s_{\rm sw}$  below which one uses the ALEPH data
to compute the integral, and above which one extrapolates  the DV Ansatz of Eq.~(\ref{DVpar}) with parameters obtained from a fit to
$V$ and $A$. To be concrete, we take the parameter values of a fit to $V$ and $A$ sum rules constructed from the three weight functions
of Eq.~(\ref{moments}) within CIPT, shown in Tab.~V of Ref.~\cite{alphas2015}. (The results are very similar if another fit set-up is chosen.) We find, for the sum rules of Eq.~(\ref{WSR1}) and~(\ref{WSR2}), the results shown
in Fig.~\ref{WSRfig} as a function of the point $s_{\rm sw}$.

\begin{figure}[ht]
\begin{center}
\subfigure[First Weinberg sum rule, Eq.~(\ref{WSR1}), without  DVs.]{\includegraphics[width=.9\columnwidth,angle=0]{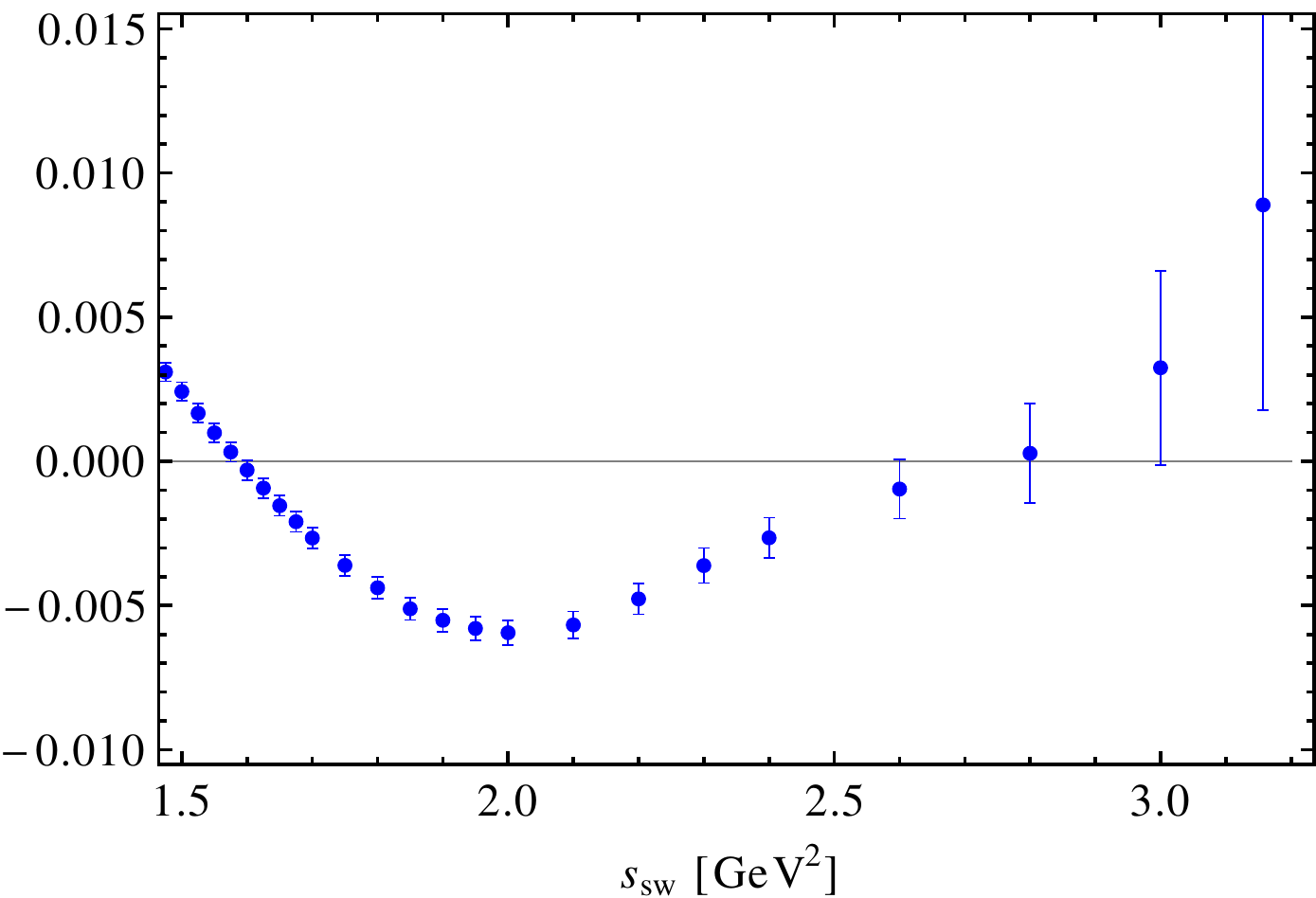}}
\subfigure[Second Weinberg sum rule, Eq.~(\ref{WSR2}), without  DVs.]{\includegraphics[width=.9\columnwidth,angle=0]{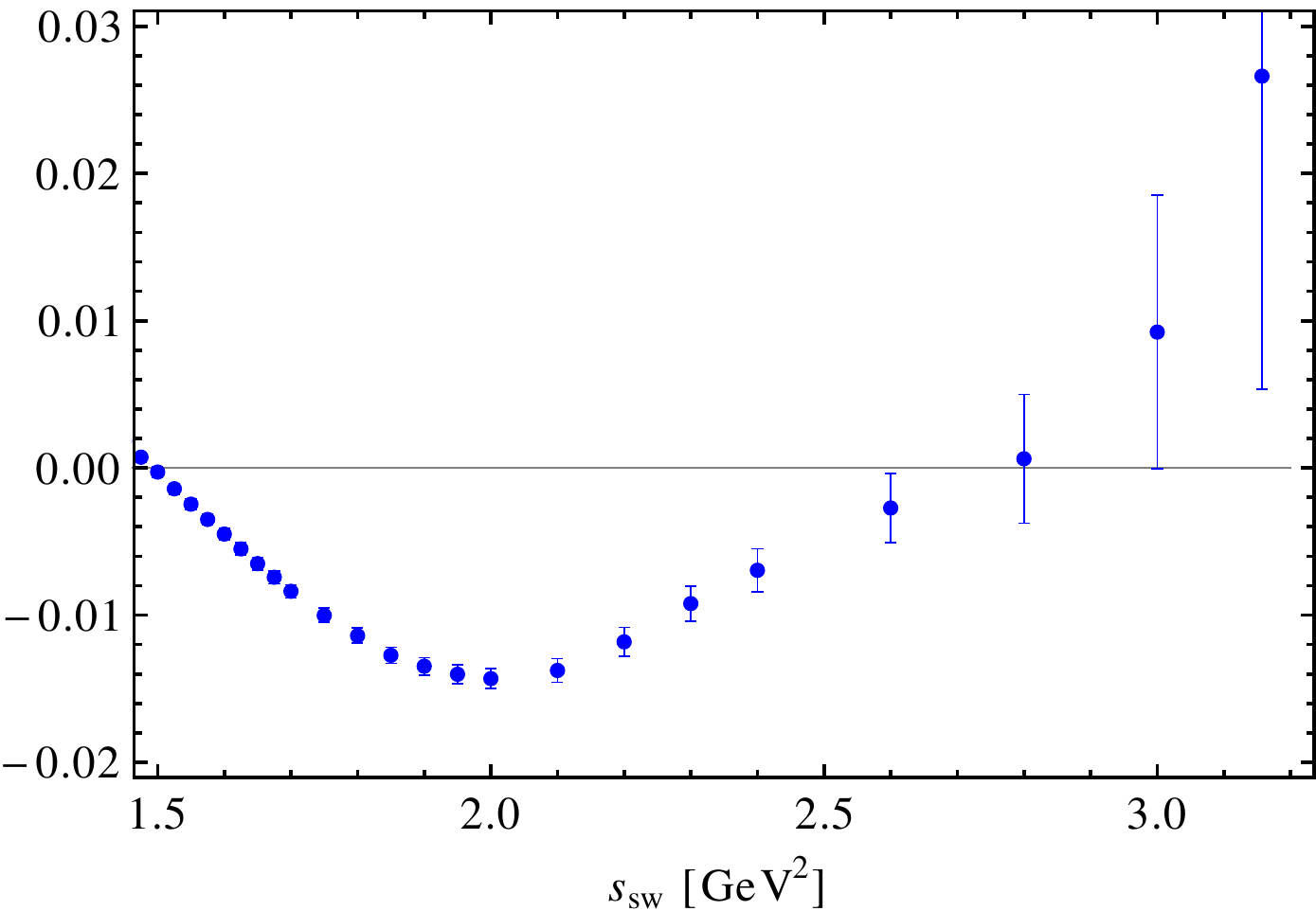}}
\caption{Weinberg sum rules, Eqs.~(\ref{WSR1}) and~(\ref{WSR2}) without the DV contribuiton. These results should be compared with Fig.~\ref{WSRfig} where the DV contribution is taken into account.}\vspace{-0.8cm}
\label{WSRnoDVs}
\end{center}     
\end{figure}

To illustrate the importance of the DVs we display in Fig.~\ref{WSRnoDVs} the results for the Weinberg sum rules as a function of $s_{\rm sw}$  but 
without the DV contribution. The comparison between these two results gives us confidence that the DVs obtained from the data are sound since their extrapolation
to infinity  is in excellent agreement with the constraints from the Weinberg sum rules.

\section{Conclusions}

We have extracted $\alpha_s$ and non-perturbative parameters from the updated ALEPH data for hadronic $\tau$ decays. This extraction
is sound and our results  fulfill consistency tests.  The $\alpha_s$ values can be averaged with those obtained from the OPAL data. The final results
are given in Sec.~\ref{alphas}. The ALEPH data constrain the DV parameters better than OPAL's and our DV parameters fulfill the constraints imposed by Weinberg sum rules.  Nevertheless, the non-perturbative physics limits the accuracy with
which the strong coupling can be extracted.

\vspace{-0.5cm}

\section*{Acknowledgements}\vspace{-0.3cm}
We would like to thank the organizers of this fruitful meeting. The work of
DB was supported by the S\~ao Paulo Research Foundation (FAPESP) grant 14/50683-0.
MG is supported in part by the
U.S. Department of Energy, and JO is supported by the U.S. Department
of Energy under Contract No. DE-FG02-95ER40896. SP is
supported by Grants No. CICYT-FEDER-FPA2014-55613-P 
and No. 2014 SGR 1450, and the Spanish Consolider-
Ingenio 2010 Program CPAN (Grant No. CSD2007-00042).
KM is supported by a grant from the Natural Sciences and
Engineering Research Council of Canada.

\end{document}